# Global fossil carbon emissions rebound near pre-COVID-19 levels

RB Jackson[1], P Friedlingstein[2,3], C Le Quere[4], S Abernethy[1], RM Andrew[5], JG Canadell[6], P Ciais[7], SJ Davis[8], Zhu Deng[9], Zhu Liu[9], GP Peters[5]

1 Department of Earth System Science, Woods Institute for the Environment, and Precourt Institute for Energy, Stanford University, Stanford, CA 94305-2210, USA

2 College of Engineering, Mathematics and Physical Sciences, University of Exeter, Exeter EX4 4QF, UK

3 Laboratoire de Météorologie Dynamique, Institut Pierre-Simon Laplace, CNRS-ENS-UPMC-X, Département de Géosciences, Ecole Normale Supérieure, 24 rue Lhomond, 75005 Paris, France

4 Tyndall Centre for Climate Change Research, School of Environmental Sciences, University of East Anglia, Norwich Research Park, Norwich, NR4 7TJ, UK

5 CICERO Center for International Climate Research, P.O. Box 1129 Blindern, 0318 Oslo, Norway

6 Global Carbon Project, CSIRO Oceans and Atmosphere, Canberra, ACT 2601, Australia

7 Laboratoire des Sciences du Climat et de l'Environnement, LSCE/IPSL, CEA-CNRS-UVSQ

8 Department of Earth System Science, University of California at Irvine, 92697 USA

9 Department of Earth System Science, Tsinghua University, Beijing 100084, China

ORCID iDs: RB Jackson https://orcid.org/0000-0001-8846-7147
P Friedlingstein https://orcid.org/0000-0003-3309-4739
C Le Quéré https://orcid.org/0000-0003-2319-0452
S Abernethy https://orcid.org/**0000-0002-3565-7243**
RM Andrew https://orcid.org/0000-0001-8590-6431
JG Canadell https://orcid.org/0000-0002-8788-3218
GP Peters https://orcid.org/0000-0001-7889-8568
P Ciais https://orcid.org/0000-0001-8560-4943
SJ Davis https://orcid.org/ 0000-0002-9338-0844
Z Liu https://orcid.org/**0000-0002-8968-7050**

*Abstract*

The years 2020 and 2021 revealed unprecedented disruptions to global economic activity and fossil carbon dioxide ($CO_2$) emissions attributable to the world's responses to the COVID-19 pandemic. Global fossil $CO_2$ emissions in 2020 decreased 5.4%, from 36.7 Gt $CO_2$ in 2019 to 34.8 Gt $CO_2$ in 2020, an unprecedented decline of ~1.9 Gt $CO_2$. We project that global fossil $CO_2$ emissions in 2021 will rebound 4.9% (4.1% to 5.7%) compared to 2020 to 36.4±0.3 Gt $CO_2$, returning nearly to 2019 emission levels of 36.7 Gt $CO_2$. Emissions in China are expected to be 7% higher in 2021 than in 2019 (reaching 11.1 Gt $CO_2$) and only slightly higher in India (a 3% increase in 2021 relative to 2019, and reaching 2.7 Gt $CO_2$). In contrast, projected 2021 emissions in the United States (5.1 Gt $CO_2$), European Union (2.8 Gt $CO_2$), and rest of the world (14.8 Gt $CO_2$, in aggregate) remain below 2019 levels. For fuels, $CO_2$ emissions from coal in 2021 are expected to rebound above 2019 levels to 14.7 Gt $CO_2$, primarily because of increased coal use in China, and will remain only slightly (0.8%) below their previous peak in 2014. Emissions from natural gas use should also rise above 2019 levels in 2021, continuing a steady trend of rising gas use that dates back at least sixty years. Only $CO_2$ emissions from oil remain well below 2019 levels in 2021. Emissions in the power and industry sectors have increased global fossil $CO_2$ emissions the most in 2021, with emissions from surface transport and aviation still below 2019 levels. The rapid rebound in global fossil $CO_2$ emissions as economies recover from the COVID-19 pandemic reinforces the need for immediate and global coherence in the world's response to climate change.

*One-line Summary*

Fossil $CO_2$ emissions in 2021 are projected to grow 4.9% (range of 4.1% to 5.7%) to 36.4±0.3 billion metric tons, pushing global emissions back close to 2019 levels (36.7 Gt $CO_2$).

*One-line Summary (<100 characters)*

Fossil $CO_2$ emissions in 2021 will grow 4.9% to 36.4±0.3 billion metric tons, returning global emissions close to 2019 levels (36.7 Gt $CO_2$).

*Main Text*

Prior to the emergence of COVID-19, average global growth in fossil $CO_2$ emissions had slowed to 0.9% annually during the 2010s (2010-2019), with global emissions in 2019 about the same as those in 2018 (~37.7 Gt $CO_2$ in both years; Friedlingstein et al. 2020, 2021). Much of the decadal slowdown in emissions growth was attributable to the substitution of coal with gas and renewables in the electricity sector (Friedlingstein et al. 2019, Jackson et al. 2016, 2019, Peters et al. 2020), and induced in part by the growing numbers of climate change laws worldwide (Eskander and Fankhauser 2020). Compared to the 2010s, average annual growth of global fossil $CO_2$ emissions was 3.0% in the 2000s, 0.9% in the 1990s, 1.6% in the 1980s, and 3.2% in the 1970s (Friedlingstein et al. 2020, 2021).

Confinement measures in response to the COVID-19 pandemic reduced social and global economic activity and $CO_2$ emissions substantially (Le Quéré et al. 2020, 2021, Liu et al. 2020, Forster et al. 2020, Friedlingstein et al., 2020, Diffenbaugh et al. 2020, IEA 2021). At the time of peak confinement in a given country, emissions decreased by one quarter (26%) on average (Le Quéré et al. 2020). Daily global fossil $CO_2$ emissions decreased 17% at peak confinement in April of 2020 (compared to 2019), and daily emissions decreased up to 75% in aviation, 50% in road transportation, and 35% in industry (Le Quéré et al. 2020). Almost half of the decline in total annual fossil $CO_2$ emissions in 2020 was attributable to reductions in transport activity (Le Quéré et al. 2020, Liu et al. 2020).

The economic disruption of COVID-19 in 2020 altered emissions in ways that varied by country, sector, and fuel type and that may have accelerated the transition to renewables. China was among the few large countries whose emissions increased in 2020 compared with 2019, despite a large but brief drop attributable to COVID-19 (Figure 1). The increase in China's total emissions was attributable primarily to its power and industry sectors, where emissions increased by ~54 and 156 Mt $CO_2$, respectively, in 2020 compared to 2019, according to preliminary estimates (Liu et al. 2020). Most of this increase took place after

April 2020 in the more industrialized maritime provinces of China (Zheng et al. 2020). Almost all other sectors and countries showed substantial reductions in $CO_2$ emissions from 2019 to 2020 (Figure 1) (Friedlingstein et al. 2021).

For fuels globally in 2020, coal use fell 6.2 EJ to 151.4 EJ/yr, a 4% decline compared to consumption in 2019 (Figure 2). Petroleum consumption decreased even more (9.6%) in 2020—an 18.2 EJ drop to 173.7 EJ/yr. Gas consumption fell a modest 2.1% to 137.6 EJ. In contrast, wind, solar, and other renewable sources jumped 10% in 2020 to 31.7 EJ, despite a substantial 25 EJ decline in global energy demand attributable to COVID-19 (Figure 2).

For 2021, we project that global fossil $CO_2$ emissions compared with 2020 levels will rebound by 4.9% (4.1% to 5.7%) to 36.4±0.3 Gt $CO_2$, nearly overlapping 2019 emission levels of 36.7 Gt $CO_2$ (Figure 1) (Friedlingstein et al. 2021). $CO_2$ emissions in 2021 are expected to rise compared to 2020 in every country and region. Our 2021 fossil $CO_2$ emissions projections are based on energy data for coal, oil and gas for the first 7 to 9 months of the year for China, USA, EU27+UK, and India, and a GDP-based projection for the Rest of the World. Full details are provided in Friedlingstein et al. (2021).

Projected fossil emissions for China in 2021 are 11.1 Gt $CO_2$, an increase of 4.0% (range 2.1% to 5.8%) compared with 2020 emissions and almost 7% higher than in 2019 (Figure 1). In fact, the largest increases across sectors and countries in 2021 compared with 2019 are found in China's power and industrial sectors (385 and 303 Mt $CO_2$, respectively; Figure 3). Projected fossil $CO_2$ emissions for India in 2021 are 2.7 Gt CO2, a substantial rebound of 12.6% (10.7% to 13.6%) compared with 2020, and slightly (~3%) above its 2019 emissions (Figure 1). In contrast, projected fossil $CO_2$ emissions for the United States and European Union in 2021 are expected to remain below 2019 levels, despite substantial increases relative to 2020 of 7.6% (5.3% to 10.0%) and 7.6% (5.6% to 9.5%), respectively, in 2021 (Figure 1). Our 2021 projections reflect long-term background trends of increasing $CO_2$ emissions for India and decreasing $CO_2$ emissions for the United States and European Union. For China, in contrast, COVID-19 recovery may have sparked growth in $CO_2$ emissions, whereas for the Rest of the World (in aggregate), it may act to dampen the recent growth in emissions (Figure 1).

For fuels in 2021, we project that $CO_2$ emissions from coal use will rebound above 2019 levels to 14.7 Gt $CO_2$ (Figure 1), primarily because of increased coal use in China (Figure 4), and will remain only slightly (0.8%) below their peak in 2014 (Figure 1). $CO_2$ emissions from natural gas use in 2021 (7.7 Gt $CO_2$) are also expected to rebound above 2019 levels; of all fossil fuels, gas use has risen steadily for at least sixty years (Figure 2). Only $CO_2$ emissions from oil remain well below 2019 levels in 2021 at an estimated 11.5 Gt $CO_2$ (Figures 1 and 2).

The distribution of the 2021 rebound in fossil $CO_2$ emissions is heterogenous across countries and sectors (Figures 1 and 3). Beyond the increases in China's power and industrial sectors in 2021 discussed above, other sectors that have also surpassed 2019 levels include power in India and Brazil (49 and 35 Mt $CO_2$ higher, respectively), residential emissions in the European Union (22 Mt $CO_2$ higher), and all sectors in Russia other than international aviation (totaling 35 Mt $CO_2$ higher) (Figure 3). These increases are balanced by sustained reductions in many other sectors, primarily international aviation emissions, which are still well below 2019 levels in all major countries (Figure 3).

Rapidly increased market penetration of renewables that displace fossil fuels is critical for limiting climate change in the 1.5° to 2°C range (Figure 2). Although most 1.5°C mitigation scenarios (e.g., van Vuuren et al. 2018) require the substitution to no- or low-carbon sources for almost all energy infrastructure by 2050, this transition is not currently occurring quickly enough to limit warming to 1.5°C (IPCC 2018). Global gas use is rising particularly quickly. Despite the temporary effects of COVID-19 to suppress energy demand and supply, gas use and its associated $CO_2$ emissions rose more than 2% a year on average for the five-year period of 2015-2020 (Figures 1 and 2). In consequence, fossil $CO_2$ emissions associated with gas use over the next few years are likely to surpass 8 Gt $CO_2$/yr. The continuing rise in natural gas use is also problematic for climate because the $CO_2$ emissions come with large and poorly constrained additional warming from methane leakage associated with fossil extraction and use (e.g., Hmiel et al. 2020, IEA 2021). Just as for coal and oil, which were also rising quickly prior to COVID-19 (Figures 1 and 2), carbon emissions from global gas use must drop rapidly if global temperatures are to stabilize below increased thresholds of 1.5 or 2 °C (Davis et al. 2019).

Climate change was revealed in many ways in 2021. Average global temperatures for the period 2017 through 2021 are expected to be between 1.1 and 1.3 °C higher than in pre-industrial times (WMO 2021). Human-induced climate change is already increasing the frequency and intensity of weather and climate extremes in virtually every region of the globe (IPCC 2021). Moreover, five years after the Paris Agreement, the emissions gap continues to grow: global emissions need to be 15 billion tons CO2e lower (for all greenhouse gases, not just $CO_2$) than current nationally determined contributions (NDCs) for a 2 °C goal, and 32 billion tons CO2e lower for the 1.5 °C goal (WMO 2021). Progress in reducing emissions is occurring, albeit slowly (Le Quéré et al. 2019, 2021, Eskander and Fankhauser 2020). Fossil $CO_2$ emissions significantly decreased in 23 countries during the decade 2010 through 2019; for the 5-year period of 2015 through 2019, fossil $CO_2$ emissions decreased in 64 countries globally (Friedlingstein et al. 2021).

The rapid rebound in global fossil $CO_2$ emissions in 2021 (returning very close to 2019 levels) is driven primarily by emissions in the power and industry sectors (Figures 3 and 5). Fossil-based investments in economic stimulus packages in post-COVID recovery plans around the world appear to have overwhelmed substantial investments in green infrastructure (Hepburn et al. 2020). The full effect of responses to the COVID-19 pandemic on CO2 emissions is still uncertain, but a further rise in emissions in 2022 cannot be ruled out—given the surface transport and aviation sectors have not yet fully recovered (Figures 3 and 5). Green investments could still work to alter underlying emissions trends, as many will take years before showing their full effects (Andrijevic et al. 2020). These trends reinforce the need for strong and globally concerted actions to slow fossil-based investments (that continue to push $CO_2$ emissions up) and to set global emissions on a trajectory consistent with the temperature limits set in the Paris Agreement.

**Acknowledgements**

The data that support the findings of this study are openly available at globalcarbonproject.org. The authors acknowledge support from the Gordon and Betty Moore Foundation (RBJ and JGC), the Australian National Environmental Science Programme's Climate Systems Hub (JGC), the European Commission Horizon 2020 projects VERIFY (#776810) (GPP, RMA, and CLQ) 4C (# 821003) (PF, GPP, RMA, CLQ), and CoCO2 (#958927) (GPA and RMA), and Future Earth. CLQ acknowledges support from the Royal Society ((project no. RP\R1\191063). We thank the many scientists and funding agencies whose efforts and support contributed to the Global Carbon Budget 2021 released by the Global Carbon Project (globalcarbonproject.org).

Figure 1. Fossil $CO_2$ emissions globally for 1990 through 2020 (top panel), with projections for 2021 shown in red in each panel, and fossil $CO_2$ emissions by fuel type (middle panel) and country or region (bottom panel) for 1959 through 2020 (Friedlingstein et al. 2021).

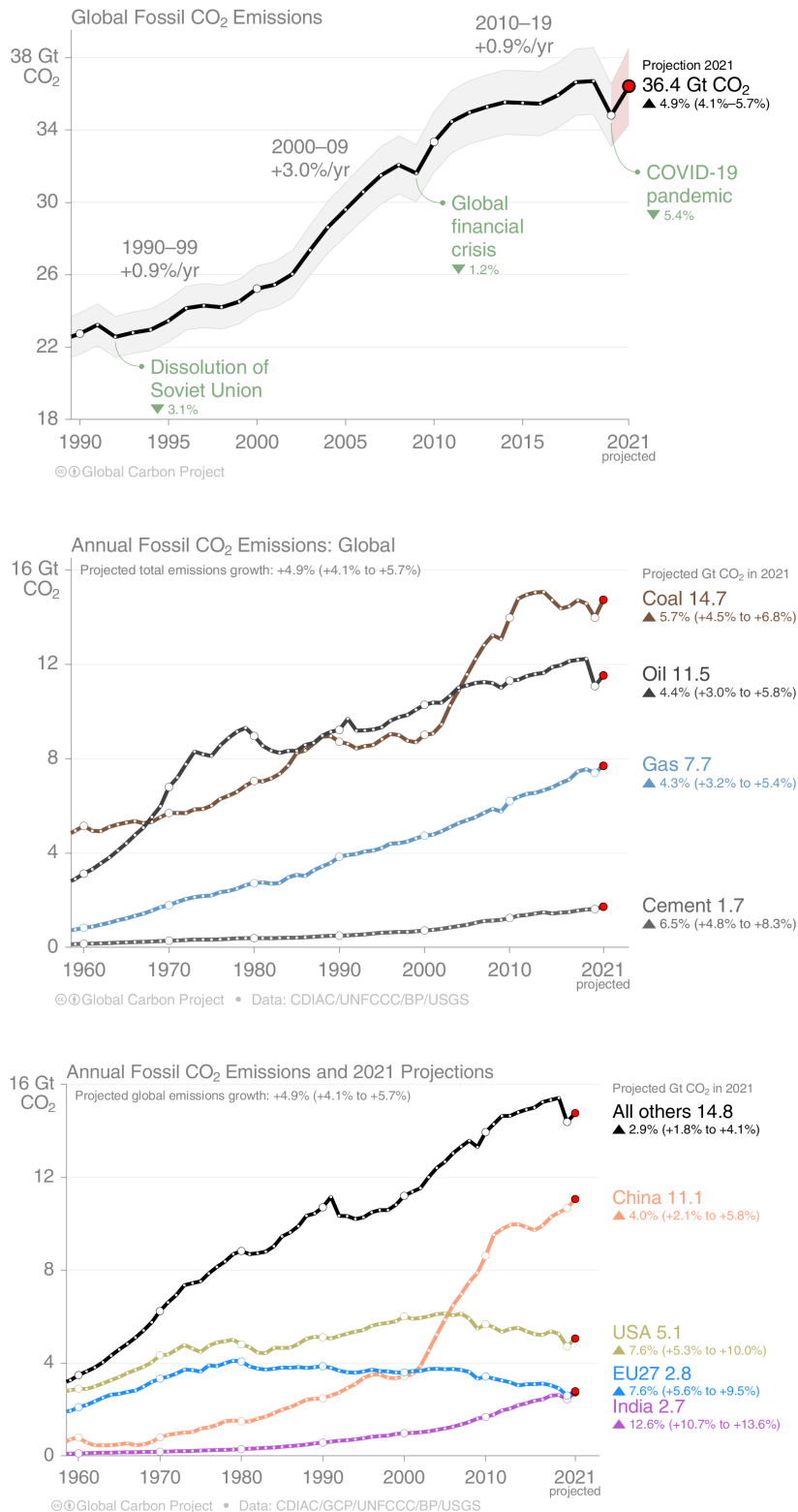

Figure 2. Top panel: Annual global energy consumption (EJ) by fuel source from 2000 through 2020 (BP 2021), with average annual growth shown under brackets for the period 2015 through 2020. Bottom panel: Fossil $CO_2$ emissions by fuel type (coal, oil, and natural gas) plus emissions from cement production and flaring; note that these emissions estimates do not include methane leakage during extraction and use.

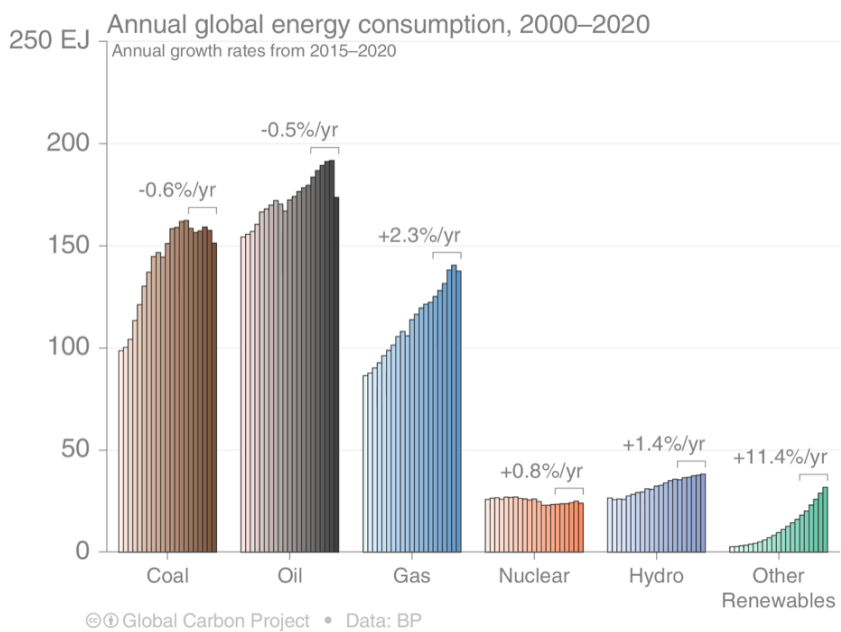

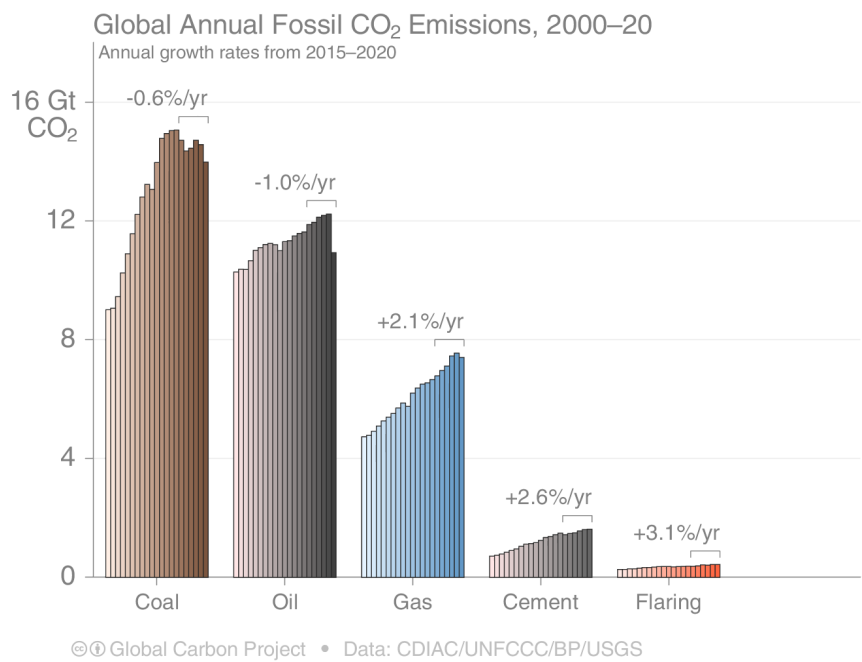

Figure 3. Fossil $CO_2$ emissions by sector and country for the difference between 2021 and 2019 emissions (Mt $CO_2$; January through September) (data from Carbon Monitor, as described in Liu et al. 2020).

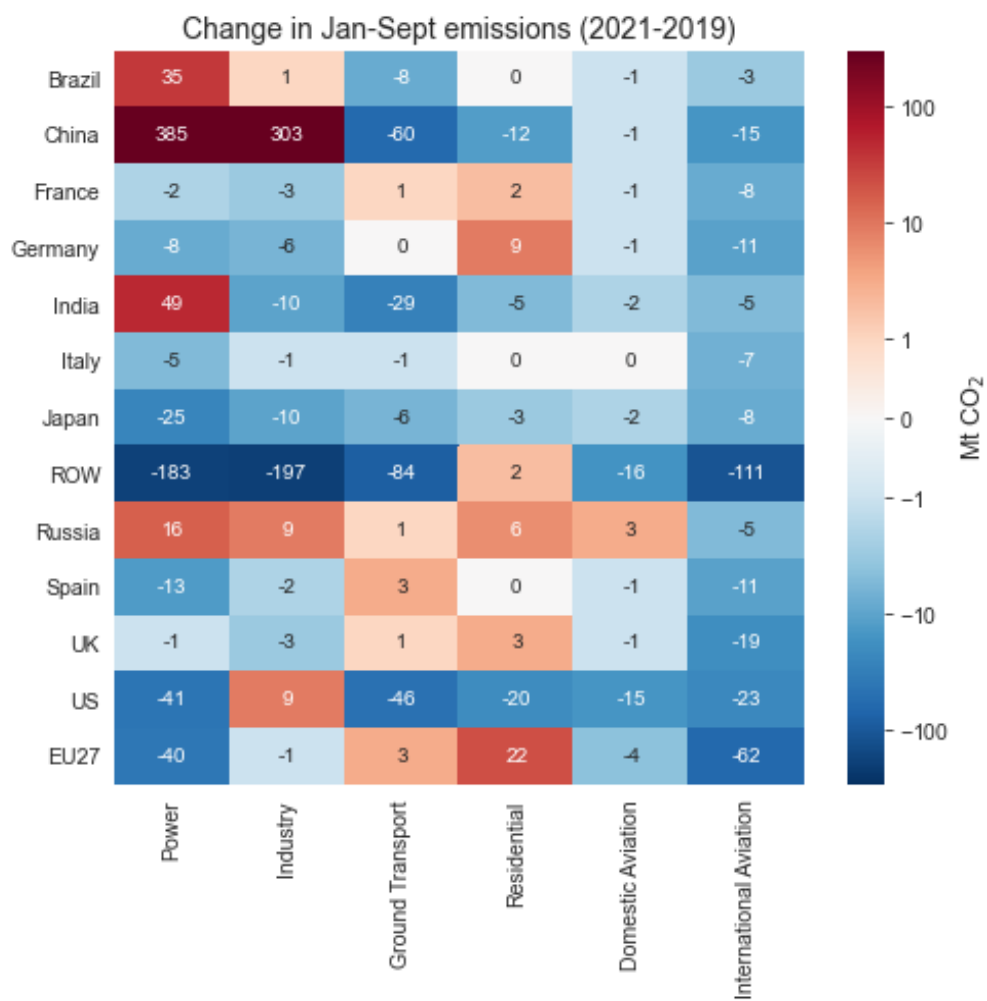

Figure 4. Annual fossil $CO_2$ emissions (Gt $CO_2$) in China by fuel type and industry. The projected emissions growth numbers for 2021 (in %) are relative to 2020 emissions.

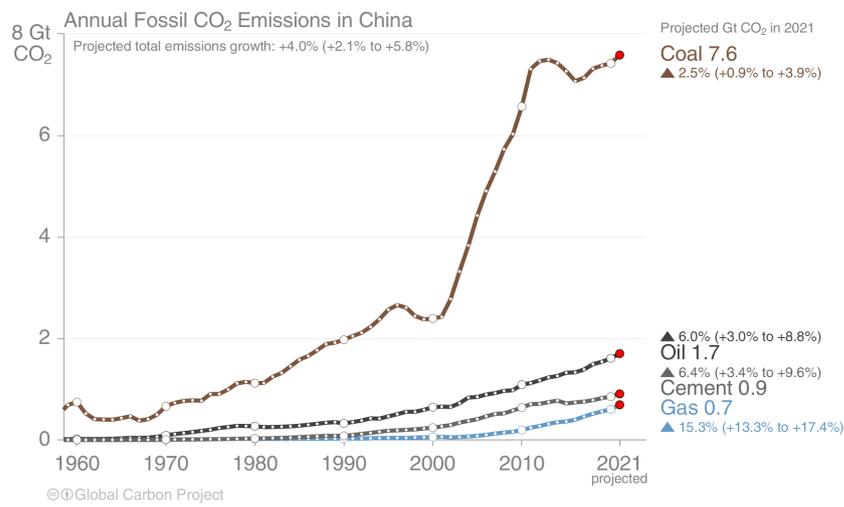

Figure 5. Daily global fossil $CO_2$ emissions for the power, industry, ground transport, and residential sectors for January through September 2021 relative to 2019 (data from Carbon Monitor, as described in Liu et al. 2020).

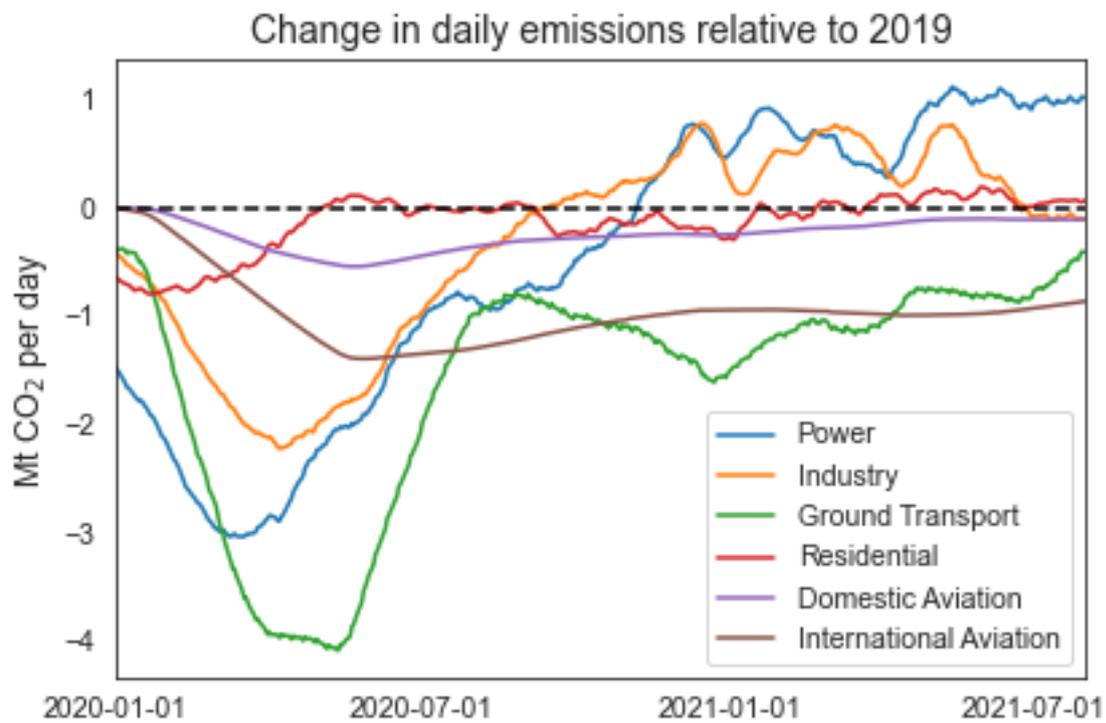